\documentclass[runningheads,a4paper]{llncs}

\usepackage{amssymb}
\usepackage{graphicx}

\usepackage{url}
\urldef{\mailsa}\path|{amelizarov, alik.kirillovich, elipachev, onevzoro}@gmail.com|
\newcommand{\keywords}[1]{\par\addvspace\baselineskip
\noindent\keywordname\enspace\ignorespaces#1}

\begin{document}

\mainmatter  % start of an individual contribution

\title{OntoMath Digital Ecosystem: \\ Ontologies, Mathematical Knowledge Analytics \\ and Management}
\titlerunning{OntoMath Digital Ecosystem}
\author{${}^1$Alexander Elizarov \and ${}^1$Alexander  Kirillovich \and ${}^1$Evgeny Lipachev\and\\
  ${}^2$Olga Nevzorova}
\authorrunning{OntoMath Digital Ecosystem}
\institute{ ${}^1$Kazan (Volga Region) Federal University, 18 Kremlyovskaya ul. Kazan, \\ 420008, Russian Federation,\\
 ${}^2$Research Institute of Applied Semiotics of Tatarstan Academy of Sciences, \\ 36à Levo-Bulachnaya ul. Kazan, 420111, Russian Federation\\
\mailsa\\
\url{http://kpfu.ru/; http://www.ips.antat.ru}}

\toctitle{OntoMath Digital Ecosystem}
\tocauthor{Elizarov, Kirillovich, Lipachev, Nevzorova}
\maketitle

\begin{abstract}
In this article we consider the basic ideas, approaches and results of developing of mathematical knowledge management technologies based on ontologies.
These solutions form the basis of a specialized digital ecosystem OntoMath
which consists of the ontology of the logical structure of mathematical documents Mocassin and ontology of mathematical knowledge OntoMathPRO,
tools of text analysis,  recommender system and other applications to manage mathematical knowledge.
The studies are in according to the ideas of creating a distributed system of interconnected repositories
of digitized versions of mathematical documents
and project to create  a World Digital Mathematical Library.
\end{abstract}

\section{Introduction}\label{intro}

The Digital Era has changed crucially as the methods of research,
and the ways in which scientists search, produce, publish, and disseminate their scientific work.
At the present time information and communication technologies are actively implemented in research and development.
Therefore, it became possible to use the entire corpus of accumulated scientific knowledge in conducting new research.
Such use requires creation of complex of technologies that ensure optimal management of available knowledge, the organization
has effective access to this knowledge, as well as sharing and multiple use of new kinds of knowledge structures.
In mathematics also accumulated considerable experience in using of electronic mathematical content within the various projects
on creation of mathematical digital libraries (see, e.g., \cite{0}).

Since inception of the first scientific information systems, mathematicians have been involved in the full cycle of software product
development, from idea to implementation. Well-known examples are an open source system \TeX{} \cite{1} and commercial systems
Wolfram Mathematica
and WolframAlpha, led by Stephen Wolfram according to his principles of computational knowledge theory \cite{2}, \cite{3}.
Tools for mathematical content management are developed with the help of communities of mathematicians,
e.g. MathJax by American Mathematical Society, information system Math-Net.Ru is developed at the Steklov Mathematical Institute of the
Russian Academy of Sciences \cite{4}--\cite{6}
and the collection of publicly available preprints arXiv.org (\url{https://arxiv.org/}).

Main challenges of mathematical knowledge management (MKM) are discussed in \cite{7}--\cite{10}. In \cite{11} we frame the most urgent tasks:
modeling representations of mathematical knowledge; presentation formats; authoring languages and tools;
creating repositories of formalized mathematics, and mathematical digital libraries; mathematical search and retrieval;
implementing math assistants, tutoring and assessment systems; developing collaboration tools for mathematics; creating new tools
for detecting repurposing material, including plagiarism of others' work and self-plagiarism; creation of interactive documents;
developing deduction systems. The solution of this task requires formalization of mathematical statements and proofs.
While mathematics is full of formalisms, there is currently none of widely accepted formalisms for computer mathematics.

At the present time one of the largest formal mathematical libraries is Mizar (\url{http://www.mizar.org/}),
which is a collection of papers prepared in the Mizar system of formal language), containing definitions, theorems and proofs \cite{12}, \cite{13}.
Mizar is one of the pioneering systems for mathematics formalization, which still has an active user community.
The project has been in constant development since 1973.

Note the important results related to the level of formalization of representations of mathematical articles.
For these purposes, developed languages of presentation of mathematical texts, specialized formal languages,
as well as conversion software languages \cite{14}--\cite{18}. These technologies are also used to construct the mathematical ontology
and creating semantic search service \cite{8}, \cite{11}, \cite{19}.
Effective communication requires a conceptualization as well as the sharable vocabulary. Ontologies suffice this requirement
\cite{8}, \cite{11}, \cite{20}.

In mathematical content important part of the search service is to find the fragments of formulas. For example,
such a service is implemented in a digital repository Lobachevskii Journal of Mathematics (\url{http://ljm.kpfu.ru/}).
For this we use converting documents in XML-format and formulas -- in of MathML-notation \cite{14}.
The above and many other mathematical implemented projects paved the way for the realization of a new idea ---
the creation of the World Digital Mathematical Library (WDML).

At the present time it formed a special type of information system called ``digital ecosystem'' \cite{21}.
The Digital Ecosystem is forming as the Information Technology, Telecommunications, and Media and Entertainment industries converge,
users evolve from mere consumers to active participants, and governments face policy and regulatory challenges \cite{22}.
Research on digital ecosystems model adaptation to scientific and educational fields are described in \cite{23}--\cite{26}.

This paper is devoted the development of the digital ecosystem OntoMath, whose task is the mathematical knowledge management
in digital scientific collections. This ecosystem is a semantic publishing platform, which forms the semantic representation
for the collections of mathematical articles and the set of ontologies and mathematical knowledge management services.

The paper is organized as follows. In Section~2, we consider the problems related to the management of digital mathematical libraries content.
In Section 3 we present the object paradigm of representation of mathematical knowledge.
Then we present the architecture and the tools OntoMath ecosystem.

\section{Digital Mathematics Libraries}\label{dml}

At present, research activities in the field of mathematics
associated with the use of modern information technology (cloud,
semantic, etc.). These technologies are used in research of
distributed scientific teams, the preparation and dissemination of
mathematical knowledge in an electronic form, the formation of
mathematical digital libraries and of intellectual processing of
their content. Special attention is given to the creation of a
common information space by mathematical integration of existing and
organizing new digital mathematical library (DML) (see, e.g., \cite{27}--\cite{30}). The largest
projects are Math-Net.RU (\cite{4}--\cite{6}, \cite{31}),
CEDRAM~\cite{32}, DML-CZ~\cite{33}, \cite{34}, DML-PL~\cite{35},
GDZ~\cite{36}, NUMDAM~\cite{37}, Zentralblatt MATH~\cite{38}, EMIS ELibM~\cite{39}, BulDML~\cite{40}. Mathematical content is presented in a
multidisciplinary digital libraries, for example, scholarly journals archives JSTOR~\cite{41} and eLIBRARY~\cite{42}.

This class also includes information of scientific publishing platform Elsewier (\url{https://www.elsevier.com/}),
Springer (\url{http://link.springer.com/}), Plei\-ades Publishing (\url{http://pleiades.online/ru/publishers/}),
as well as system support of scientific journals, for example, Elpub (\url{http://elpub.ru/}).

Realization and development of digital mathematics libraries involve the development of special tools and continuous improvement of
their functionality. An example is the Open Journal Systems (OJS)~\cite{43}--\cite{45}.
The platform used in many projects, particularly in Lobachevskii Journal of Mathematics (\url{http://ljm.kpfu.ru/}),
one of the first digital mathematical journals. In the practice of this journal, since 1998,
been introduced intelligent information processing tools \cite{14}, \cite{46}--\cite{49},
in particular, we performed automated MathML-markup articles of this journal \cite{14}, \cite{49}.
In paper \cite{50} presents a system of services for the automated processing of large collections of scientific documents.
These services provide verification of document compliance to the accepted rules of formation of collections and their
conversion to the established formats; structural analysis of documents and extraction of metadata,
as well as their integration into the scientific information space.
The system allows to automatically perform a set of operations that cannot be realized
for acceptable time with the traditional manual processing of electronic content.
It is designed for the large collections of scientific documents.

The idea of creating a World Digital Mathematical Library (WDML) arose in 2002.
The initial aim of this project was digitizing the entire set of mathematical literature (both modern and historical) and link it to the present literature,
and make it clickable (see~\cite{27}, \cite{51}--\cite{54}). As noted in \cite{52}, the success of this project and its future impact on mathematics,
science and education could be the most significant event since the invention of scientific journals and to become a prototype for a new model of scientific
and technical cooperation, a new paradigm for the future of science electronically connected world.
At the same time, the implementation of such a large project will inevitably cause a series of problems.
These problems and ways to overcome them were analyzed in~\cite{55}. In particular, one of the recommendations was the proposal to develop and coordinate some
local projects of creating DML (see~\cite{28}, \cite{55}).

Basic plans for the construction of WDML in 2014--2015 discussed various mathematical communities and enshrined in a number of documents (see~\cite{56}, \cite{57}).
In particular, it was noted that the next step in the development of the project WDML will building information networks, knowledge-based,
contained in mathematical publications. In the discussion of these ideas was attended by many research groups of mathematicians all over the world,
including our group of Kazan Federal University. In February 2016 in the Fields Institute (Toronto, Ontario) by the Wolfram Foundation, the Fields Institute,
and the IMU/CEIC working group for the creation of a World Digital Mathematics Library it was organized the Semantic Representation of Mathematical
Knowledge Workshop~\cite{58}.
Our report on this symposium was devoted to the modeling and software solutions in the area of semantic representation of mathematical knowledge~\cite{59}.
These results correspond to the general ideology WDML project of part semantic representation and processing of mathematical knowledge and are a strategic
direction of research of our group. In particular, they are connected with the construction OntoMath ecosystem, which is described below.

\section{Moving Towards the Object Paradigm of Mathematical Knowledge Representation}\label{object}

E-libraries as a collection of electronic documents provide a document search by their bibliographic descriptions and thematic classification codes,
as well as full-text search within the documents by keywords.
Creating a full text index is a main mechanism for text search.

The global initiative WDML specifies key areas related to both organizational efforts of the international mathematical community,
including mathematical literature publishers, and research and technology efforts aimed at development of existing and introduction of new (semantic)
technologies of representation and processing of mathematical content.
These semantic technologies include the following:
\begin{itemize}
    \item Aggregation of different ontologies, indexes, and other resources created by the mathematical community,
    and ensuring broad access to their replenishment and editing;
    \item Improving the access to mathematical publications --
    not only to searching and browsing, but for annotating, navigation, linking to other sources, data computing, data visualization, and so on.
\end{itemize}
The move towards the representation of the internal structure of mathematical knowledge creates a new paradigm of representation.
The focus of representation has shifted to the selection of elements (classes) and their relationships, which allows researchers to create
various network conceptual frameworks (e.g. the citation graph, the graph of mathematical concepts, etc.).
Classification of mathematical objects and organization of the relevant repositories provide new computing capabilities for data processing
such as extraction and processing of formulas, finding similar papers and so on.

WDML project is focused on the object system of organization and storage of mathematical knowledge.
Unlike traditional electronic mathematical libraries in which the unit storage in the database is an electronic document,
it is proposed to provide the mathematical knowledge of the collections of documents in the form of specially organized repository of mathematical objects.

One of the key ideas is to develop the classes of objects for adequate description and study of mathematical content.
In a mathematical document it is easy enough to identify a set of basic classes of mathematical objects
(sequences, functions, transformations, identity, symbols, formulas, theorems, statements, etc.).
As noted in WDML project, one of the most important tasks is to build a list of mathematical objects in different areas of mathematics.

Standard classes of mathematical objects are theorems, axioms, proofs, mathematical definitions, etc.
Important elements of the object model are semantic links (relations) between the elements.
In order to build a document object model, it is proposed to use modern technologies of the Semantic Web.
This representation of mathematical knowledge requires development of new management tools that will be relevant to mathematical knowledge
(aggregation tools, semantic search, search formulas and identification of similar objects) \cite{53}, \cite{56}, \cite{57}.

Let us consider the key objects of mathematical knowledge management tools.

Aggregation tools provide automatic collection of objects that meet certain criteria, as well as automatic replenishment of object lists.
Object lists can be built according to different criteria, depending on the target application.
For example, a useful list is a list of objects of a given domain (for example, a list of all known theorems of the group theory),
or a list of objects of a particular class (e.g. theorem) related to the study of the mathematical properties of a given object
(e.g. the geometric object ``triangle''). These lists allow to actually creating custom search indexes which would accumulate mathematical knowledge.

Navigation tools (with search tools) provide opportunities for navigation to target objects within the document.
For example, the classical task is to find a given mathematical object and its properties, and to search for this given mathematical object and other
mathematical objects related to certain mathematical equations.
Another important task is to find the given mathematical object and scientific articles on this subject.
At the same time, in contrast to the keyword search, object search would allow to take into account the semantics of links for object search,
thereafter to improve search results.

For example, using the object properties for a given mathematical object (e.g., ``Sobolev space'') it is possible to find and view
relevant information about such properties as its mathematical definition, educational literature, context-related objects and others
(see, e.g., Section~\ref{recomm}).

Semantic search is the method of information retrieval which determines the relevance of the document to the request semantically rather than syntactically.
Semantic search in the object repository is organized by following semantic links that allow to find objects by their description (implicit reference to object),
as well as by given object properties.
For example, the following query is classified as an implicit reference to the object: ``Find all the theorems, the proof of which is used Fermat's theorem.''

Search by formulas: this search tool provides search of mathematical formulas and additional information about them (such as the name of the formula,
the list of scientific and educational publications, etc.). Formula search queries, in general, can have different forms.
For example, a text query to the variables (``Find a formula connecting the area of the circle and the length of its circumference''),
or computing request (``Find a formula equivalent to the formula, the $F$''), or text query to the mathematical object connected with this formula
(``Find evidence of Euler's formula'').

Thus, the main purpose of WDML is to unite digital versions of all mathematical repositories, including both contemporary sources and sources that have become historical,
on new conceptual base and to provide intelligent information retrieval and data processing~\cite{56}, \cite{57}.

At the same time more and more popular in the scientific community have become new ways to detect objects of scientific knowledge directly through the web,
as well as tools and services for creating and sharing of new types of knowledge structures.
In the context of the concept of Linked Data, and the Semantic Web these tools and services can be used to create ``cooperation graphs'' (collaboration graph),
which are used, for example, to calculate the collaboration distance between the authors and searching similar documents.
These facilities open up new possibilities of fine-tuning searching and browsing (see, e.g., \cite{60}).
Many authors (e.g. \cite{8}, \cite{11}, \cite{20}) highlight the importance of developing new domain ontologies, in particular in mathematics,
because the traditional bibliographic classification is no longer sufficient.
It needs a deeper representation that would contain more detailed descriptions by taking into account different points of view.

\begin{figure}
\centering
\includegraphics[width=2.5in]{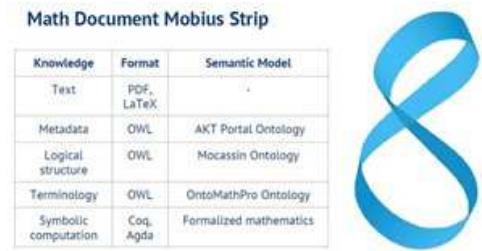}
\caption{Levels of representation of mathematical content.}
\label{fig:mobius}
\end{figure}

We proposed in [59] model of representation of mathematical content in the form of a Mobius strip (Fig.~\ref{fig:mobius}).

\section{OntoMath Digital Ecosystem}\label{ecosystem}

\subsection{General Description}\label{general}

OntoMath is a digital ecosystem of ontologies, textual analytics tools, and applications for mathematical knowledge management.
This system consists of the following components:

\begin{itemize}
    \item Mocassin, an ontology of structural elements of mathematical scholarly papers;
    \item $\mathrm{OntoMath^{PRO}}$, an ontology of mathematical knowledge concepts;
    \item Semantic publishing platform;
    \item Semantic formula search service;
    \item Recommender system.
\end{itemize}

\begin{figure}
\centering
\includegraphics[width=3.7in]{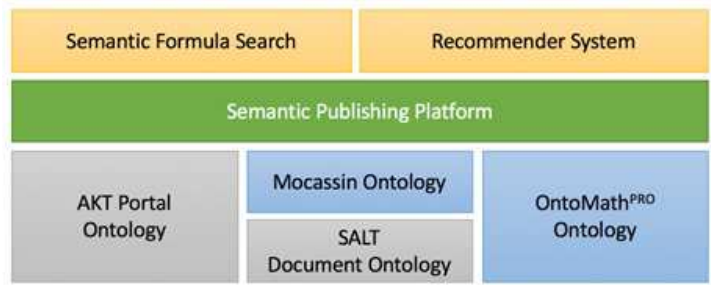}
\caption{OntoMath ecosystem architecture.}
\label{fig:architect}
\end{figure}

Briefly we describe these basic elements of the architecture of OntoMath digital ecosystem (Fig.~\ref{fig:architect}).

The core component of the OntoMath ecosystem is its semantic publishing platform. It builds an LOD representation for a collection of mathematical articles in \LaTeX{}.
The generated mathematical dataset includes metadata, the logical structure of documents, terminology, and mathematical formulas.
Article metadata, the logical structure of documents, and terminology are expressed in terms of AKT Portal, Mocassin and $\mathrm{OntoMath^{PRO}}$ ontologies respectively.
Mocassin ontology, in turn, is built on Semantically Annotated \LaTeX{} (SALT) Document Ontology that is ontology of the rhetorical structure of scholarly
publications~\cite{SALT}.
Mocassin and $\mathrm{OntoMath^{PRO}}$ ontologies are parts of OntoMath ecosystem but SALT is an external ontology. Two applications are built using the
semantic publishing platform: a semantic formula search service and a recommender system.

As any digital ecosystem, OntoMath has components that are used for socio-technical purposes. Such components are the ontologies and the semantic publishing platform.
They can be used by mathematicians and software developers.

\subsection{Semantic Publishing Platform}\label{platform}

As was mentioned above, the semantic publishing platform which constitutes the core of the OntoMath ecosystem makes an LOD representation for a given
sample of mathematical articles in \LaTeX{} \cite{61}, \cite{62}. Its main features are:

\begin{itemize}
    \item Indexing mathematical articles in \LaTeX{}-format as LOD-compatible RDF-data;
    \item Extracting articles' metadata in terms of AKT Portal Ontology~\cite{AKT};
    \item Mining the document logical structure using our ontology of structural elements of mathematical papers;
    \item Eliciting instances of mathematical entities as the concepts of $\mathrm{OntoMath^{PRO}}$ ontology;
    \item Connecting the extracted textual instances to symbolic expressions and formulas in the mathematical notation;
    \item Establishing the relationship between published data and RDF-existing sets of LOD data.
\end{itemize}

The developed technology has the following features:

\begin{itemize}
    \item Mathematics RDF-set is based on a collection of mathematical articles in Russian;
    \item The RDF-built set that includes metadata and also specific semantic knowledge such
as the knowledge generated as a result of special treatment of mathematical formulas
(binding textual definitions of variables with the symbols of variables in formulas)
and also the instances of $\mathrm{OntoMath^{PRO}}$ ontology and the structural elements of the
mathematical articles;
    \item Semantic annotation of mathematical texts based on ontologies Mocassin and $\mathrm{OntoMath^{PRO}}$;
    \item The MathLang Document Rhetorical (DRa) Ontology~\cite{64} enables to interpret the
elements of document structure using mathematical rhetorical roles that are similar
to the ones defined in the statement level of OMDoc ontology. This semantics focuses
on formalizing proof skeletons for generation proof checker templates.
\end{itemize}

\subsection{Ontologies}\label{ontologies}

Mocassin~\cite{63} is an ontology, intended to annotate a logical structure of a mathematical document~\cite{61}, \cite{65}.
This ontology extends SALT Document Ontology, defining concepts and relations, specific to mathematical documents.
Mocassin ontology represents a mathematical document as a set of interconnecting segments.

Mocassin ontology defines15 concepts such as
\textit{Document segment}, \textit{Claim}, \textit{Definition}, \textit{Proposition}, \textit{Example}, \textit{Axiom}, \textit{Theorem},
\textit{Lemma}, \textit{Proof}, \textit{Equation}, and others.
The ontology defines relations between segments such as \textit{dependsOn}, \textit{exemplifies}, \textit{hasConsequence}, \textit{hasSegment},
\textit{proves}, \textit{refersTo}.

\begin{figure}
\centering
\includegraphics[width=4in]{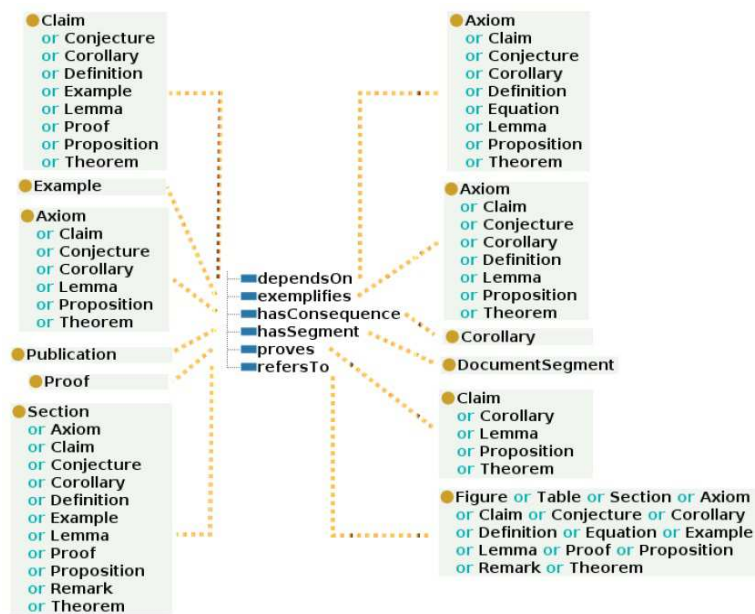}
\caption{Mocassin ontology Elements \cite{66}.}
\label{fig:mocassin}
\end{figure}

$\mathrm{OntoMath^{PRO}}$ \cite{67} is an ontology of mathematical knowledge~\cite{11}, \cite{68}, \cite{69}. Its concepts are organized into two taxonomies:

\begin{itemize}
    \item Hierarchy of areas of mathematics: \textit{Logics}, \textit{Set theory}, \textit{Geometry}, including its subfields,
    such as \textit{Differential Geometry} and so on;
    \item Hierarchy of mathematical objects such as a \textit{set}, \textit{function}, \textit{integral}, \textit{elementary event}, \textit{Lagrange polynomial} etc.
\end{itemize}

The ontology defines relations:

\begin{itemize}
    \item Taxonomic relation (for example, ``Lambda matrix'' \textit{is} a ``Matrix'');
    \item Logical dependency (for example, ``Christoffel Symbol'' \textit{is defined by} ``Connectedness'');
    \item Associative relation between objects (for example, ``Chebyshev Iterative Method'' \textit{see also} ``Numerical Solution of Linear Equation Systems'');
    \item Belongingness of objects to fields of mathematics (for example, ``Barycentric Coordinates'' \textit{belongs to} ``Metric Geometry'');
    \item Associative relation between problems and methods (for example, ``System of linear equations'' \textit{is solved by} ``Gaussian elimination method'').
\end{itemize}

\begin{figure}
\centering
\includegraphics[height=6cm]{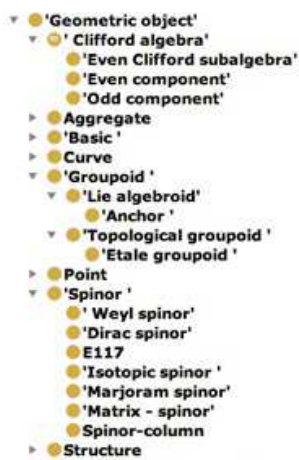}
\caption{OntoMath${}^{\mathrm{PRO}}$ ontology Elements.}
\label{fig:ontomathpro}
\end{figure}

Each concept description has Russian and English labels, textual definitions, relations with other concepts, and links to external terminologies,
such as DBpedia~\cite{DB}, \cite{DB2} and ScienceWISE~\cite{ScienceWISE}.

\subsection{Applications}\label{applicatoins}

OntoMath Formula Search Engine is a semantic search service that uses a semantic representation of math document built on the base of
Semantic Publishing Platform~\cite{62}, \cite{68}. OntoMath Formula Search Engine implements new search on names of variables using OntoMath ontology.
A variable in the formula is a symbol that denotes a mathematical object.
Mathematical symbols can designate numbers (constants), variables, operations, functions, punctuation, grouping, and other aspects of logical syntax.
Specific branches and applications of mathematics usually have specific naming conventions for variables.
However, nonstandard names of variables may be used in some formulas.
OntoMath Formula Search Engine allows finding the mathematical formulas containing a given mathematical object regardless of its name for the variable.
For example, if we would like to find a formula that contains a mathematical object (e.g. the curvature),
the service will find all the formulas that include this object (even with different names for the variable).
Using an inference, the service can find the formulas containing not only the given object, but the objects below in the hierarchy of the ontology.
For example, for searching the formulas which contain the polygon, OntoMath Formula Search can find the formulas which contain not only the polygon
but other objects in the hierarchy (e.g. the triangle, the parallelogram, the trapezium, the hexagon and others).
OntoMath formula search also allows restricting
your search to the document area that you define. For example, you can search only in the defined areas or in a certain theorem area.
These search functions of OntoMath Formula Search Engine differ from those of popular search services, such as (uni) quation,
Springer \LaTeX{} search, Wikipedia search formula, Wolfram search formula. These services have a great potential, including their stability for renaming variables
and for expression transformation.
However, they are syntactic and seek formulas containing a predetermined formula pattern.

We have implemented two applications for mathematical formula search such as syntactical search of formulas in MathML, and semantic ontology-based search.

The syntactical search leverages formula description from documents formatted in \TeX{}.
Our algorithm~\cite{14} transforms formulas from \TeX{} format to MathML format. We set up an information retrieval system prototype for a collection of articles
in Lobachevskii Journal of Mathematics. For the end-user, the query input interface supports a convenient syntax.
The search results include highlighted occurrences of formulas as well as document metadata.

\begin{figure}
\centering
\includegraphics[height=6.2cm]{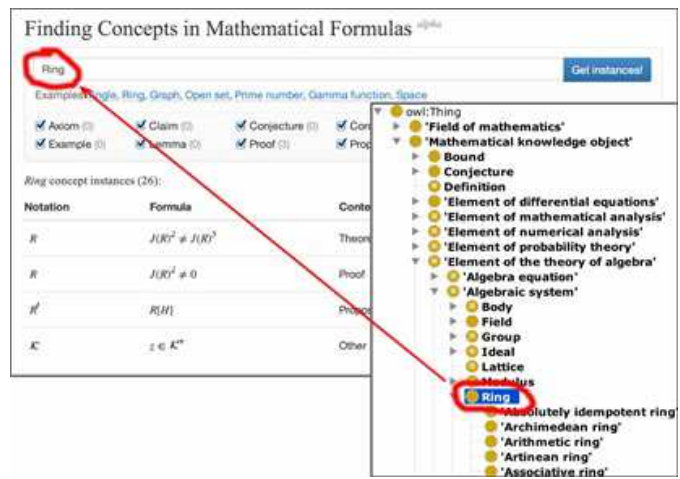}
\caption{OntoMath Search UI.}
\label{fig:search}
\end{figure}

\subsection{OntoMath Recommender System}\label{recomm}

As ecosystem OntoMath application we have developed a recommender system for collections of physical and mathematical documents.
One of the main functions of this system is the creation of the list of related documents (see~\cite{70}--\cite{72}).
Traditionally, the list of related documents is based on the keywords given by the authors, as well as bibliographic references available in the documents.

This approach has several disadvantages:

\begin{itemize}
    \item A list of keywords may be missing or incomplete;
    \item A keyword may be ambiguous;
    \item It is necessary to take into account the hierarchy of concepts;
    \item It should be noted that the article may use the terminology in different languages.
\end{itemize}

Thus, the created recommender system has the following features:

\begin{itemize}
    \item It takes into account the professional profile of a particular user;
    \item It forms different recommendations for different scenarios of work with the system (referee, user being introduced in the topic, etc.);
    \item It assigns different weights to different concepts. Thus, for a scientific review, concepts denoting areas of mathematics are more
    important than those related to mathematical objects.
    While for a beginning researcher, survey papers containing notions from different areas of mathematics and references to original works are more important.
\end{itemize}

\begin{figure}
\centering
\includegraphics[height=6.8cm]{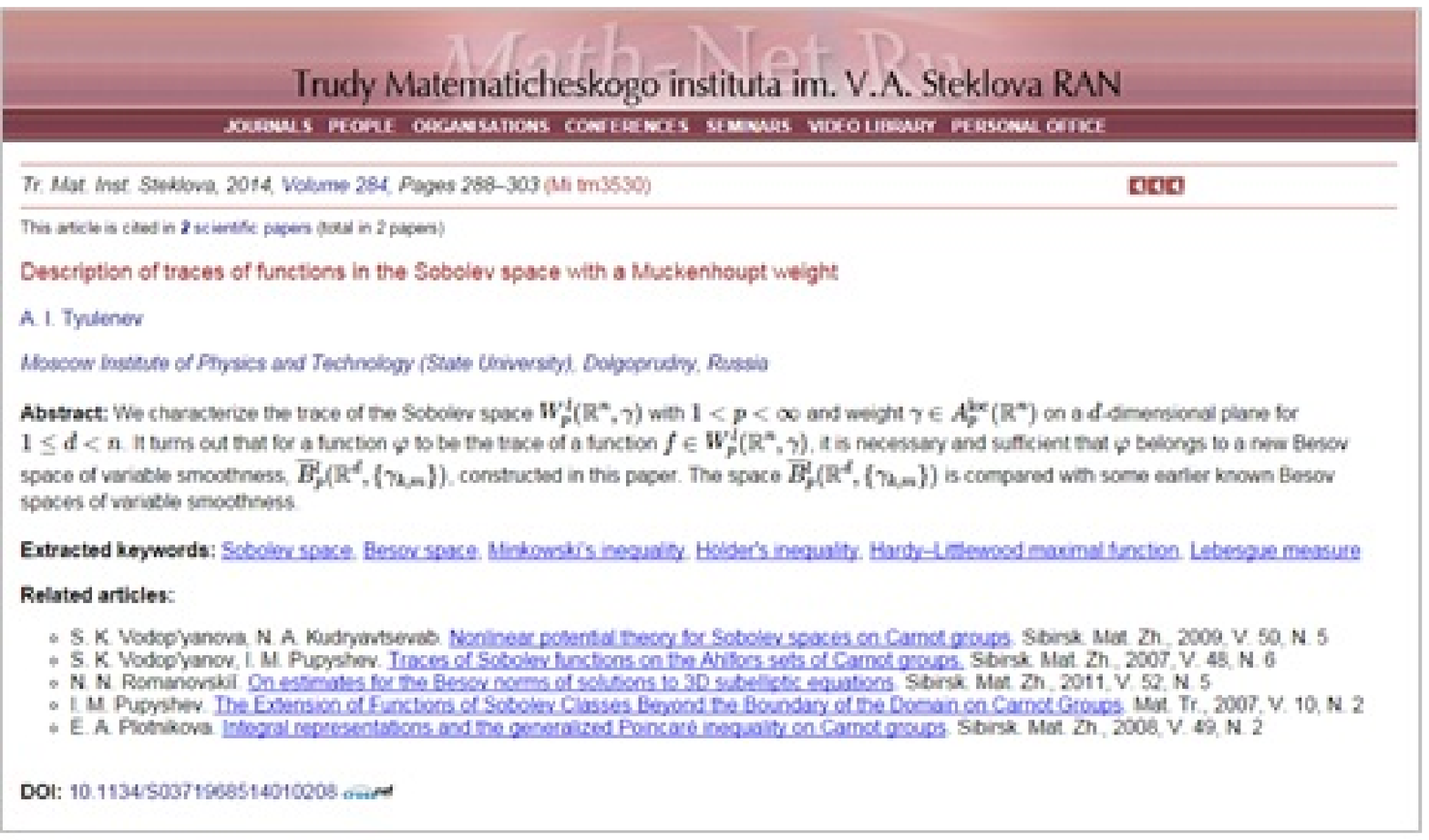}
\caption{Math-Net article page. Extracted keywords and related articles are added.}
\label{fig:card}
\end{figure}

\section{Conclusion}\label{conclusion}

The basic ideas, approaches and results of developing the discussed mathematical knowledge management technology are based on targeted ontologies
in the field of mathematics. These solutions form the basis of the specialized digital ecosystem OntoMath which consists of a set of ontologies,
text analytics tools and applications for managing mathematical knowledge.
The studies are in line with the project aimed to create a World Digital Mathematical Library whose objective is to design a distributed system
of interconnected repositories of digitized versions of mathematical documents.

The future of the OntoMath ecosystem is related to the development of new services for semantic text analytics and control of mathematical knowledge.
The developed technologies are supposed to be evaluated with the help of the digital mathematical collections of Kazan Federal University.

The present work is aimed at further research in the field of mathematical knowledge management; it has been carried out by the authors of the paper since 1998,
with the support of grants from the Russian Foundation for Basic Research, Kazan Federal University and the Academy of Sciences of the Tatarstan Republic.

\section*{Acknowledgement}

This work was funded by the subsidy allocated to Kazan Federal
University for the state assignment in the sphere of scientific
activities, grant agreement no.~1.2368.2017).

\end{document}